\documentclass[%
 reprint,
 amsmath,amssymb,
 aps,
prb,
]{revtex4-2}

\DeclareMathAlphabet\mathbfcal{OMS}{cmsy}{b}{n}

\usepackage{graphicx}
\usepackage{dcolumn}

\usepackage{bm,dsfont}

\usepackage{xcolor}

\newcommand{\comment}[1]{}

\begin{document}

\preprint{APS/123-QED}

\title{Multiphoton Spectroscopy of a Dynamical Axion Insulator}

\author{Olivia Liebman}
 \email{oliebman@ucla.edu}
\author{Jonathan B. Curtis}%
 \email{joncurtis@ucla.edu}
\affiliation{%
 College of Letters and Science, University of California, Los Angeles
}%
\author{Ioannis Petrides}%
 \email{ipetrides@g.ucla.edu}
\affiliation{%
 College of Letters and Science, University of California, Los Angeles
}%
\author{Prineha Narang}
 \email{prineha@ucla.edu}
\affiliation{%
College of Letters and Science, University of California, Los Angeles
}%

\date{\today}

\begin{abstract}
The unusual magnetoelectric transport present in Weyl semimetals and 3D topological insula- tors can be compactly understood as manifestations of a background axion field, which itself is determined by the microscopic band structure. In the presence of correlations, an additional axion quasiparticle may emerge as the collective excitations on top of the mean background field. Such modes couple nonlinearly to electric and magnetic fields, giving rise to a dynamical magnetoelectric response. However, unambiguous identification of this collective axion mode is challenging due to its inherent nonlinear dynamics. Here, we propose an all-optical protocol that utilizes a pump-probe setup for verifying and characterizing the transient dynamics of axion fields in three-dimensional insulator systems. In particular, we show that nonlinear Raman processes induce dynamical oscillations of the axion field that depend on the geometry of the incident electromagnetic fields. These oscillations manifest in the polarization and magnetization of the material, hence, can be subsequently measured using time-resolved Kerr rotation spectroscopy. Our results open a pathway towards using multi-photon and quantum pair spectroscopies to identify new correlated phases of quantum matter.

\end{abstract}

\maketitle

\noindent
\textit{Introduction}\textemdash Axionic particles were originally proposed in high-energy physics in order to solve the charge-parity problem as they were associated to the Nambu-Goldstone bosons of a new global axial U(1) symmetry~\cite{Peccei_Quinn_1977_1, Peccei_Quinn_1977_2}.  
Today, axions are a leading dark matter candidate and could help explain the matter-antimatter asymmetry present in the universe~\cite{Day_Ellis_Marsh}. 
While the axion has yet to be discovered in particle physics, its condensed matter counterpart has been theorized to exist in certain three-dimensional materials~\cite{Nenno.2020,Li.2010,Wang.2013,Roy.2015,Wan.2011,Pace.2021,petrides2022semiclassical}.
A characteristic identifier of the presence of an axion is a signature magnetoelectric response, leading to a modified form of Maxwell's equations known as axion electrodynamics~\cite{Sekine2020AxionEI,Zyuzin.2012cdp,Essin.2009};  
these give rise to physical phenomena which can broadly be classified as either transport or density responses. 
Axion-induced transport includes anomalous Hall currents which are related to the gradient of the axion, and a magnetic conductivity that relates to the axion's first time derivative. 
Density responses have to do with the topological magnetoelectric effect where the axion acts as the coupling constant between magnetic field and polarization; and similarly between  electric field and magnetization.

In correlated systems, the axion field has an associated collective excitation which couples {\it nonlinearly} to the electromagnetic field; this induces a dynamical magnetoelectric effect on top of the mean field response due to the background axion field~\cite{Srivatsa.2018,Wieder.2020,Sehayek.2020,McKay.2021}.
A variety of candidate materials have been put forward where axion field effects manifest in the electronic band structure.
For example, axion fields naturally appear in correlated Weyl semimetals that undergo lattice translation symmetry breaking due to the emergence of charge density wave (CDW) order.
In this case, the mean background axion field is determined by the separation of Weyl points in momentum space and in energy; while its dynamical contribution is associated with collective excitations of the phase of the CDW, i.e., the Nambu-Goldstone bosons~\cite{Burkov12,Shi.2021}. 
In the presence of uncompensated carrier densities the softening of the axionic collective mode may even lead to spatially inhomogeneous correlated phases~\cite{Curtis.2023}.
3D topological insulators (TI) can also support a static, background axion field which, due to time-reversal symmetry, acts as a $\mathbb{Z}_2$ topological invariant and takes quantized values $\theta = 0$ or $\pi$. 
When interactions are turned on that spontaneously break time-reversal symmetry, as in the case of antiferromagnetic (AFM) order, $\theta$ is promoted to a dynamical quantity that takes continuous values between $0$ and $2 \pi$~\cite{Li.2010}. 
Here, the fluctuations of the axion field are induced by the spin-wave excitations of the AFM order parameter~\cite{Li.2010, Wang.2020, Zhang.2020}. 

The emergence of axion physics in condensed matter systems has ignited hope of studying such elusive physics in a tangible, tabletop experimental setting. 
While the high energy axion continues to go undetected,  
measuring axion quasiparticles in condensed matter systems have posed its own theoretical and experimental challenges due its inherent nonlinear coupling. 
Prominent techniques rely on studying linear perturbations in the presence of a strong magnetic field, for example in measuring the attenuated total reflection~\cite{Li.2010}, or positive longitudinal magnetoconductance~\cite{Gooth.2019}. 
Piezomagnetic probes have also been used to identify the dynamical bulk orbital magnetization induced by strain~\cite{Yu.2021}, while others relied on detecting surface plasmons~\cite{Zhu.2022,Li.2010}.
These approaches have proven valuable for detecting equilibrium responses, however, collective dynamics often rely on nonlinearities that induce a more complex evolution.

\begin{figure*}
\includegraphics[width=0.8\textwidth]{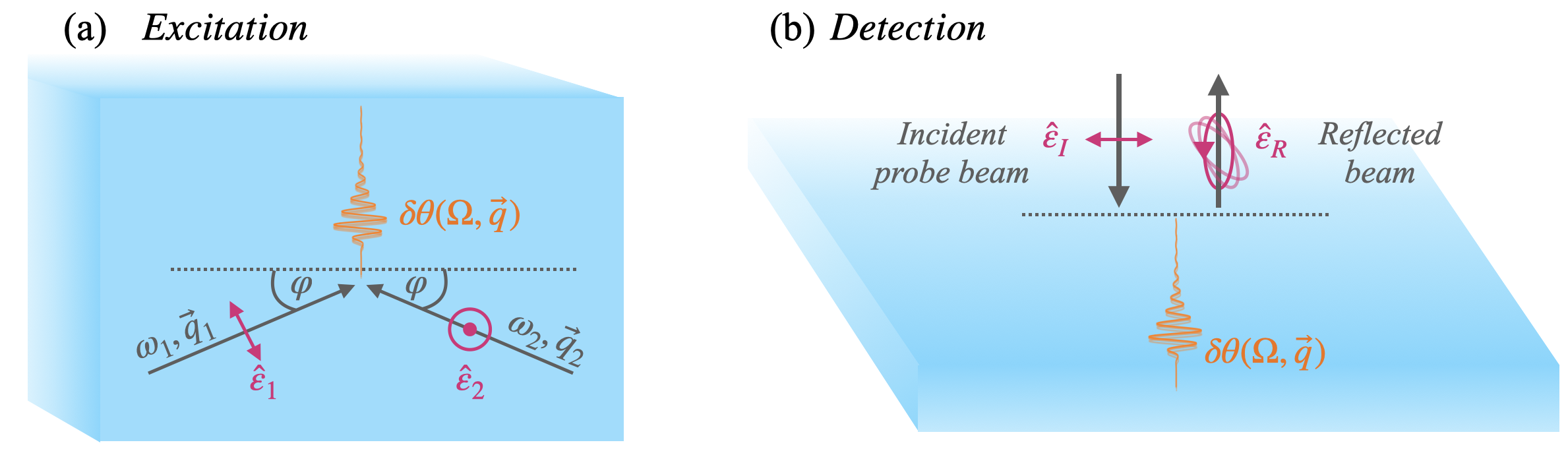}
\caption{\label{fig:schematic} Schematic of the general excitation and detection set-up (a) Excitation scheme:  two incident linearly polarized beams, $A^{(1)}(\omega_1, \boldsymbol{q}_1)$ and $ A^{(1)}(\omega_2, \boldsymbol{q}_2)$, come in at a relative angle of $\pi - 2\varphi$, with respective linear polarizations $\boldsymbol{\hat{\varepsilon}}_1,\boldsymbol{\hat{\varepsilon}}_2$; frequencies $\omega_1,\omega_2$; and momenta $\boldsymbol{q}_1, \boldsymbol{q}_2$. The two  beams are used to induce finite $\boldsymbol{E}\cdot\boldsymbol{B}$, exciting an axionic collective mode using a multiphoton absorption event. 
(b) Detection scheme: a linearly polarized probe beam, with polarization $\boldsymbol{\hat{\varepsilon}}_I$, at normal incidence; 
upon reflection the beam's polarization $\boldsymbol{\hat{\varepsilon}}_R$ will in general be elliptically polarized at a polarization angle that oscillates in phase with the induced axion oscillations.}
\end{figure*}
Here we define an all-optical, contactless protocol to identify nonlinear signatures of dynamical axion excitations in a three-dimensional insulating system, termed a \textbf{dynamical axion insulator} (DAI).
Our general approach applies to any system hosting axionic excitations.
The protocol has two parts: first, a dynamical axion mode is resonantly excited from a background axion field via two-photon Raman processes.
By solving the nonlinear coupling between electromagnetic fields and collective axion modes, we show that the axion’s transient evolution depends sensitively on the incident angle, polarization, and frequencies of the two photons.
After excitation, a third probe beam detects the resulting axion electrodynamics by monitoring Kerr angle oscillations, serving as a faithful proxy for the axion field’s dynamics~\cite{Wu.2016,Ahn.2022}.
We apply our protocol to a minimal Hamiltonian model that serves as the basis of describing a variety of candidate materials.

\textcolor{black}{Our all-optical, contactless protocol offers advantages over existing methods. Ahn et al. \cite{Ahn.2022} study static, frequency-dependent axions in magnetic TIs requiring broken time-reversal symmetry. In contrast, our method applies to general axionic materials with fluctuating collective modes. While Li et al. \cite{Li.2010} and Gooth et al. \cite{Gooth.2019} infer dynamical axions indirectly via static fields or transport, we directly track axion oscillations using time-resolved Kerr rotation—without linearizing the equations of motion—yielding a more general and sensitive probe.
}

\textit{Axion electrodynamics}\textemdash 
We start by considering the axion-modified electromagnetic Lagrangian 
\begin{eqnarray}\label{eq:maxwellaxion}
\mathcal{L}_{\rm EM} =  \frac{1}{2} \left( \epsilon \boldsymbol{E}^2 - \frac{1}{\mu} \boldsymbol{B}^2 \right) + \frac{\alpha}{\pi} \theta \boldsymbol{E} \cdot \boldsymbol{B}.
\label{eq:Lagr}
\end{eqnarray}
The first two terms are the sourceless Maxwell Lagrangian in terms of the electric (magnetic) fields $\boldsymbol{E}$ ($\boldsymbol{B}$), and $\epsilon$ ($\mu$) is the permittivity (permeability) tensor.
The last term proportional to $\boldsymbol{E}\cdot \boldsymbol{B}$ is the so-called axion response that leads to a magnetoelectric coupling mediated by the $\theta$ term, and $\alpha \approx \frac{1}{137}$ is the fine-structure constant. Note that natural units $c = \hbar =  1$ are used throughout unless otherwise specified. 

From the axion-modified electromagnetic Lagrangian~\eqref{eq:Lagr} the current density is given by
\begin{equation}\label{eq:AxionCurrent}
  \boldsymbol{J} = \frac{\alpha}{\pi} \nabla \theta \times \boldsymbol{E} + \frac{\alpha}{ \pi} 
\partial_t \theta\boldsymbol{B},
\end{equation} 
which quantifies the electromagnetic response of the system due to the axion term.
The first term in the above equation is equivalent to an anomalous Hall response proportional to the spatial gradient of the axion field, while the second term is known as the chiral magnetic effect and is only relevant in the presence of time-dependent axion fields. 

In a dynamical axion insulator, $\theta$ is a degree of freedom which has its own equations of motion and may depend on space and time, $\theta(\boldsymbol{r},t) =  \theta_0 (\boldsymbol{r},t)+  \delta\theta(\boldsymbol{r},t)$,
where the first term is the background axion, while the second term $\delta\theta(\boldsymbol{r},t)$ is the {\bf dynamical axion mode}. 
$\theta_0$ is determined from the microscopic band structure alone, whereas $\delta \theta$ is a dynamical field with distinct equations of motion that couple nonlinearly to the electromagnetic fields.

When this is the case, the electromagnetic Lagrangian of Eq.~\eqref{eq:Lagr} is supplemented by the equations of motion of a spinless bosonic field that characterizes the dynamical axion mode, namely
\begin{eqnarray}\label{eq:FullLagrangianLorentzForm}
     \mathcal{L}_{\rm NG} = \frac{\kappa}{2} \left[ \left(\partial_t \delta\theta\right)^2  - \Omega_0^2  \delta\theta^2 \right], 
\end{eqnarray}
where $\kappa$ is a materials-dependent parameter, to be given a physical interpretation as ``compressibility" later on when we consider the specific case of an axion insulator. 
We have also included a finite gap for the collective mode $\sim \Omega_0$, which reflects the possibility of a resonance at finite frequency, e.g., due to collective mode pinning from disorder. 

As we are dealing with electrodynamics, we assume that the speed at which the axion propagates is orders of magnitude slower than the speed of light; as such the axion is effectively dispersionless provided it has a finite gap $\Omega_0$. 
Comparing to the speed of light, $c \gg \nu$ so that 
all terms relating to first gradients of the dynamical axion field, $\nabla \delta \theta$, can be safely ignored; these describe effects from axion dispersion through the medium. 

The equation of motion for the dynamical axion mode $\delta\theta$ are given by  
\begin{equation}\label{eq:EOM_theta}
     \kappa \left(\partial_t^2 + \gamma \partial_t + \Omega_0^2  \right)\delta \theta = \frac{\alpha}{\pi} \boldsymbol{E}\cdot\boldsymbol{B},
\end{equation} 
where we have included a phenomenologically relevant dissipation term $\kappa \gamma \partial_t \delta \theta$, with damping rate $\gamma$, which represents the coupling of the order parameter to other degrees of freedom in the bulk. 
The above equation describes the dynamics of a massive spinless bosonic field, driven by an effective force $\frac{\alpha}{\pi} \boldsymbol{E}\cdot \boldsymbol{B}$. 

\textit{Dynamical axion generation}\textemdash 
Exciting a dynamical axion mode in a material is a nonlinear optical process that minimally requires interactions with two electromagnetic fields, see Eq.~\eqref{eq:EOM_theta}. 
As such we consider the following plane-wave ansatz $\boldsymbol{A}^{(0)}(q) =\mathbfcal{A}_1 \delta_{q,q_1} + \mathbfcal{A}_2 \delta_{q,q_2}$, composed of two waves with distinct frequency and momentum, whose field amplitudes $\mathbfcal{A}_i = \mathcal{A}_i \boldsymbol{\hat{\varepsilon}}_i$ are related to the polarizations $\boldsymbol{\hat{\varepsilon}}_i$ which obey $\boldsymbol{q}_i \cdot \boldsymbol{\hat{\varepsilon}}_i = 0$. 
The response of the dynamical axion at first order $\delta \theta = \alpha \delta \theta^{(1)} + O(\alpha^2)$  is given by (see S.I. for a detailed derivation)
\begin{eqnarray}\label{eq:main_theta_soln}
       \delta \theta(q) = \frac{(\alpha /\pi \kappa ) \omega_1 \omega_2\mathcal{A}_1  \mathcal{A}_2  }{\Omega^2 + i \Omega \gamma - \Omega_0^2}   \left( 
\frac{\boldsymbol{q}_1}{\omega_1}  -  \frac{\boldsymbol{q}_2}{\omega_2} \right) \cdot \left(\boldsymbol{\hat{\varepsilon}_1} \times \boldsymbol{\hat{\varepsilon}_2} \right), \label{eq:theta_soln} 
\end{eqnarray}
where axion frequency is $\Omega = \omega_1 + \omega_2$ and axion momentum $\boldsymbol{q} = \boldsymbol{q}_1 + \boldsymbol{q}_2$ due to energy-momentum conservation.
In general, the intensity of the axion response depends on the frequency, momenta, polarization, and angle of the incident beams, see Fig.~\ref{fig:schematic}(a).

\begin{figure}[t]
\includegraphics[width=0.4\textwidth]{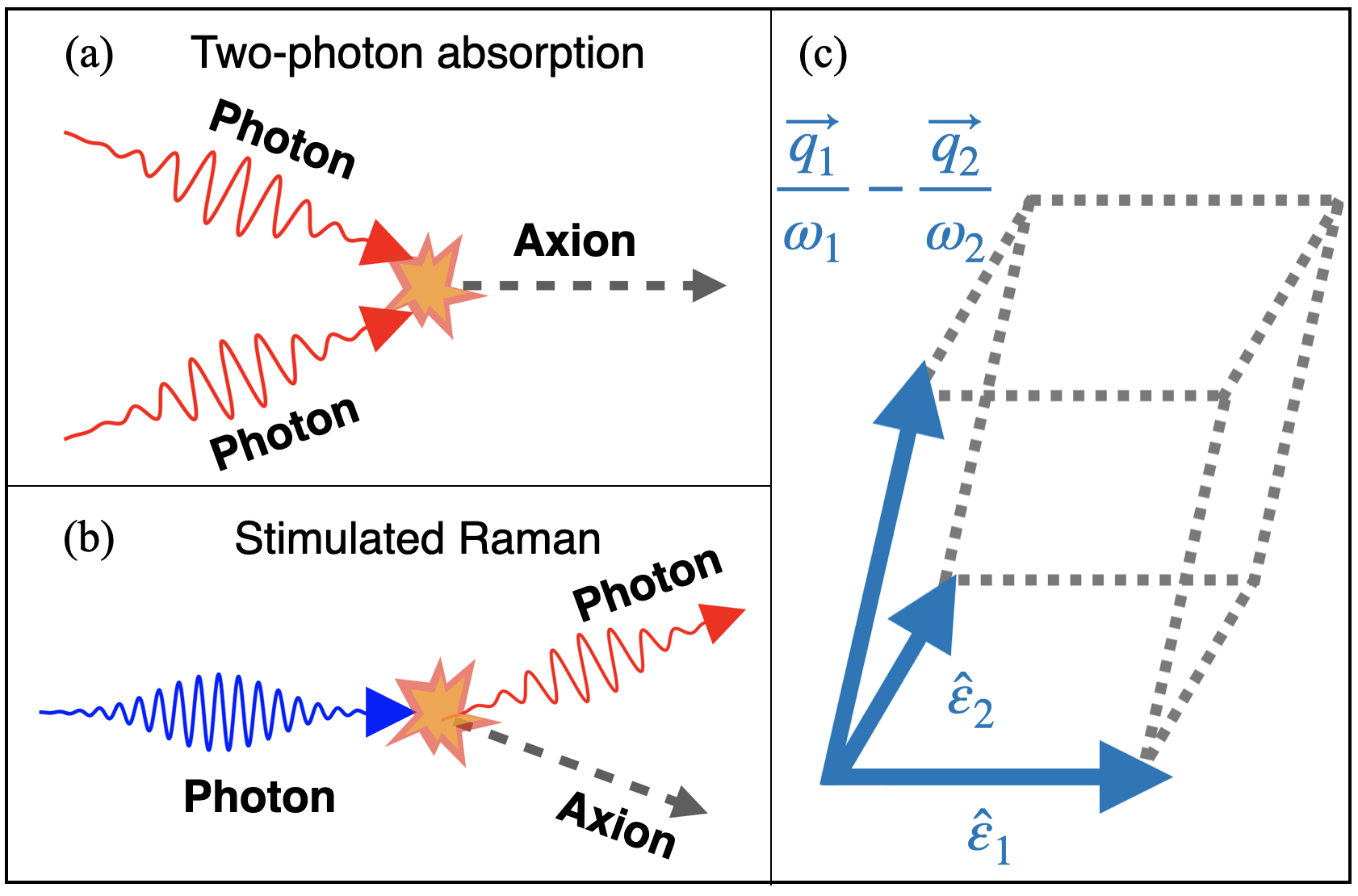}
\caption{\label{fig:TwoPhoton} Schematic depiction of different axion excitation pathways.
(a) Two photons can be absorbed in order to produce an axion excitation, which is most efficient when $\omega_1,\omega_2 \sim \Omega_0/2$.
(b) Stimulated Raman can excite the axion when an incoming photon inelastically scatters and emits an axion collective mode, leaving the sample with a different energy. (c) Parallelpiped volume generated by triple product in Eq.~\eqref{eq:main_theta_soln}.}
\end{figure}

Importantly, the axion oscillation amplitude in Eq.~\eqref{eq:theta_soln} depends on: (i) a geometric volume set by polarization vectors $\hat{\boldsymbol{\varepsilon}}_i$ and momenta $q_i / \omega_i$, and (ii) a frequency-dependent Lorentzian response. As a second-order nonlinear optical process, excitation occurs via two channels: two-photon absorption (sum or second harmonic generation), Fig.\ref{fig:TwoPhoton}(a), and stimulated Raman scattering (difference frequency generation), Fig.\ref{fig:TwoPhoton}(b). The channel dependence arises from the parallelepiped volume formed by $\hat{\varepsilon}_i$ and $\boldsymbol{q}_1 / \omega_1 - \boldsymbol{q}_2 / \omega_2$, as shown in Fig.\ref{fig:TwoPhoton}(c), constraining the response by beam geometry.

\begin{figure*}
\includegraphics[width=0.8\textwidth]{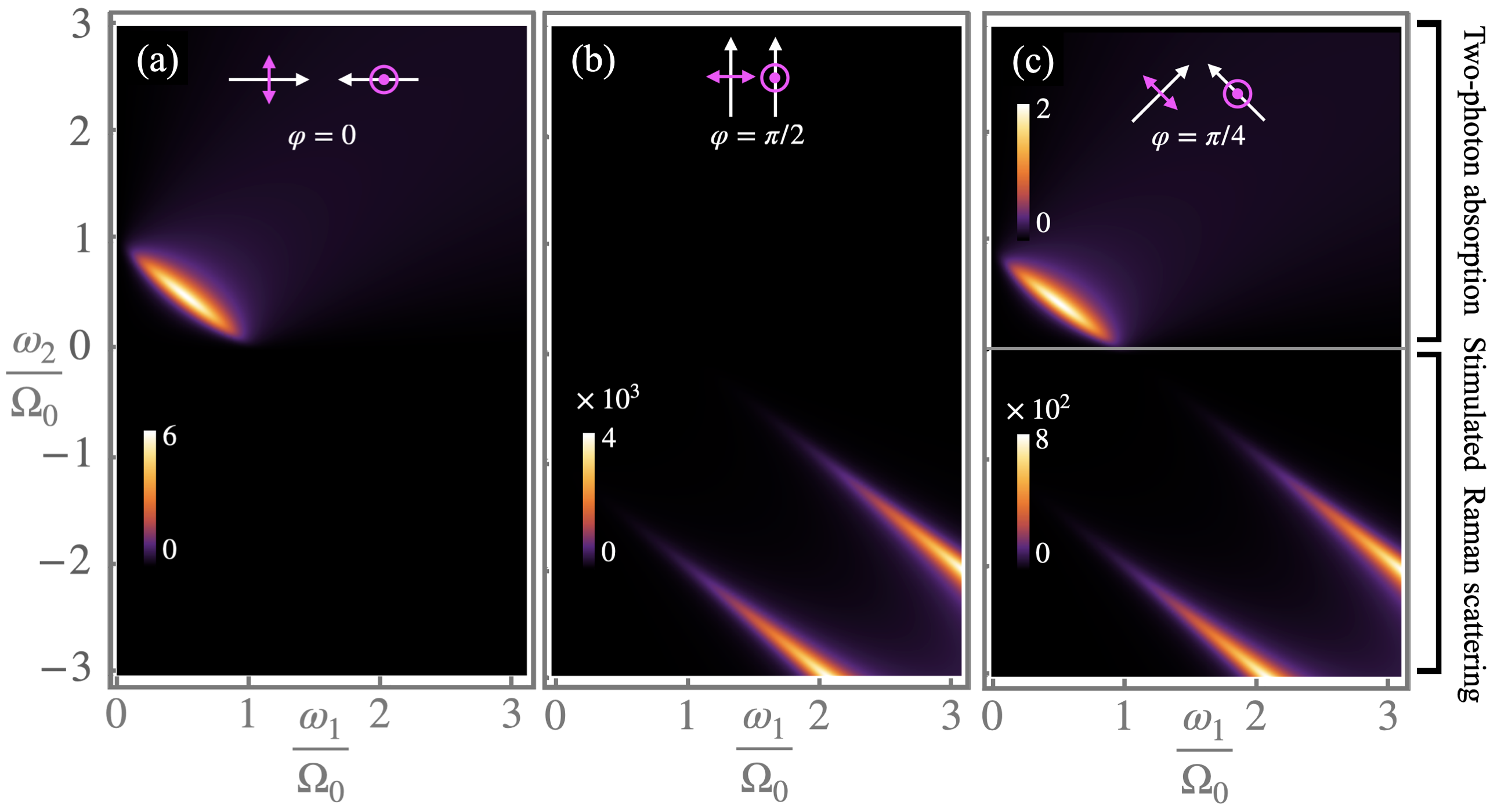}
\caption{\label{fig:plots1} 
This figure illustrates two distinct axion excitation pathways: two-photon absorption (upper half-plane) and stimulated Raman scattering (lower half-plane), with excitation amplitude $|\delta \theta|^2$ shown via color bars. The amplitude is plotted against normalized beam frequencies $\omega_1 / \Omega_0$ and $\omega_2 / \Omega_0$, with $\Omega_0 = 1$, and beam geometries indicated atop each subfigure.
(a) For $\varphi = 0$ (anti-parallel beams), two-photon absorption dominates.
(b) At $\varphi = \pi/2$ (parallel incidence), two-photon absorption is suppressed, and Raman excitation is maximal.
(c) For $\varphi = \pi/4$ (orthogonal beams), both channels are present, with Raman excitation strongly enhanced. The grey line at $y = 0$ separates the two excitation regimes.}
\end{figure*}

The efficacy of this two-photon excitation protocol for a variety of different incident beam angles $\varphi$ and frequencies $\omega_i$ in the maximally-crossed polarization geometry, i.e., we keep the polarizations of the excitation beams orthogonal $\hat{\varepsilon}_1 \cdot \hat{\varepsilon}_2 = 0$, c.f., Fig.~\ref{fig:schematic}.  
In Fig.~\ref{fig:plots1}, we show the induced dynamical axion amplitude $|\delta \theta|^2$ as a function of incident photon frequencies $\omega_1, \omega_2$ for different incident angles, $\varphi = 0, \pi/4$, and $\pi/2$. 
Fig.~\ref{fig:plots1} clearly shows how varying degrees of freedom of the incident beam can excite the axion field, with the inelastic stimulated Raman channel for collinear beams resulting in the strongest response. 

\textit{Axion detection}\textemdash
We now move on to demonstrate how the axion collective mode may be detected. The key idea is to use a linearly polarized detection beam of known frequency at normal incidence, which when reflected off the surface of the material will posses a time-dependent Kerr effect induced by the presence of the dynamical axion ~\cite{Wu.2016}. 

The proposed pump-probe style protocol is laid out in Fig.~\ref{fig:schematic}, where the pump beam excites the dynamical axion mode in the bulk and is subsequently probed through the vacuum by the reflected field.  
The background axion in the material is assumed to break both inversion and time-reversal symmetry, such that it gives rise to a bulk Hall conductivity $\nabla\theta_0$, and bulk magnetic conductivity $\partial_t{\theta}_0$, respectively [c.f. Eq.~\eqref{eq:AxionCurrent}]. 
We note that in the presence of time-reversal  symmetry the Hall effect, related to $\nabla \theta_0 $, will vanish; correspondingly, in the presence of inversion symmetry the chiral magnetic effect, related to $\partial_t \theta_0 $, will vanish.

The dispersion relation of electromagnetic fields in the material is obtained by solving the equations of motion of the electromagnetic field $  \epsilon \partial^2 _t{\bm A}- \frac{1}{\mu}\nabla^2 \bm A =\partial_t \bm J$, 
where $\epsilon$ is the relative permittivity, $\mu$ is the relative permeability, and $\bm J$ is the current given in Eq.~\eqref{eq:AxionCurrent}.
Without loss of generality we assume a non-magnetic, spatially isotropic medium such that the permeability and permittivity tensors are simplified to $\mu^{-1} =\mathds{1}$ and $\epsilon =\epsilon \mathds{1}$, where $\epsilon$ is a constant characterizing the material. 
Furthermore, we consider propagation along the z-axis and solve the equations of motion in the plane orthogonal to the direction of propagation. 
The refractive index given by (see S.I.)
$  n_{\pm} = \left( \frac{\epsilon \pm \epsilon'}{1\pm \mu' } \right)^{1/2} $,
with corresponding dispersion $k  = \omega n_{\pm}$,
where $\epsilon' = \omega^{-1}{\partial_z \theta}$ is related to the electric conductivity and $\mu' =k^{-1}{\partial_t \theta}$ is related to the magnetic conductivity.

The reflectance and transmittance amplitudes can be obtained via the continuity of electric and total magnetic fields at the interface, given as  $\boldsymbol{E}^I +\boldsymbol{E}^R = \boldsymbol{E}^T$, and $\boldsymbol{B}^I + \boldsymbol{B}^R = \boldsymbol{B}^T + \frac{\alpha}{\pi} \delta\theta \boldsymbol{E}^T$ (see S.I.).  
Without loss of generality, here we only include the dynamical axion $\delta \theta$ contributions to the Fresnel equations, and set the background axion at the boundary to zero. 
Any such contributions, as in $\mathds{Z}_2$ topological insulators where the axion takes quantized values, can be incorporated in the equilibrium position of the dynamical axion mode $\delta\theta$.
Using the matching conditions, the coefficients of reflection are computed as
\begin{equation}
\label{eqn:Rpm_main}
    R_{\pm} = \pm \frac{1 - (n_{\pm} \mp i \frac{\alpha}{\pi} \delta \theta  )}{1 + (n_{\pm} \mp i \frac{\alpha}{\pi} 
\delta \theta )}  , 
\end{equation}
where $R_{\pm} = \frac{A_{\pm}^R}{A_{\pm}^I}$ is the ratio between the reflected and initial amplitudes of the fields.
The reflection coefficients are, in general, different depending on the refractive index of left- and right-circularly polarized light, and the dynamical axion angle.  

By linearizing Eq.~\eqref{eqn:Rpm_main} for small $\delta \theta $ one obtains the following expression for the polar Kerr rotation angle:
\begin{align}\label{eq:Kerr}
   \Theta_K \approx  \frac{\alpha}{\pi} \frac{(-2 + n_+^2 + n_-^2 )}{(n_+^2 -1)(n_-^2 -1)} \delta \theta .
\end{align}
The final result of this work is contained within the above equation, which shows the modulation of the reflected beam's polar Kerr rotation angle and its dependence on the collective fluctuations of the axion field, c.f., Eq.~\eqref{eq:theta_soln}. This serves as our smoking gun optical signature to unambiguously deduce the presence of a dynamical axion field in the material.

\begin{figure}[b]
\includegraphics[width=0.3\textwidth]{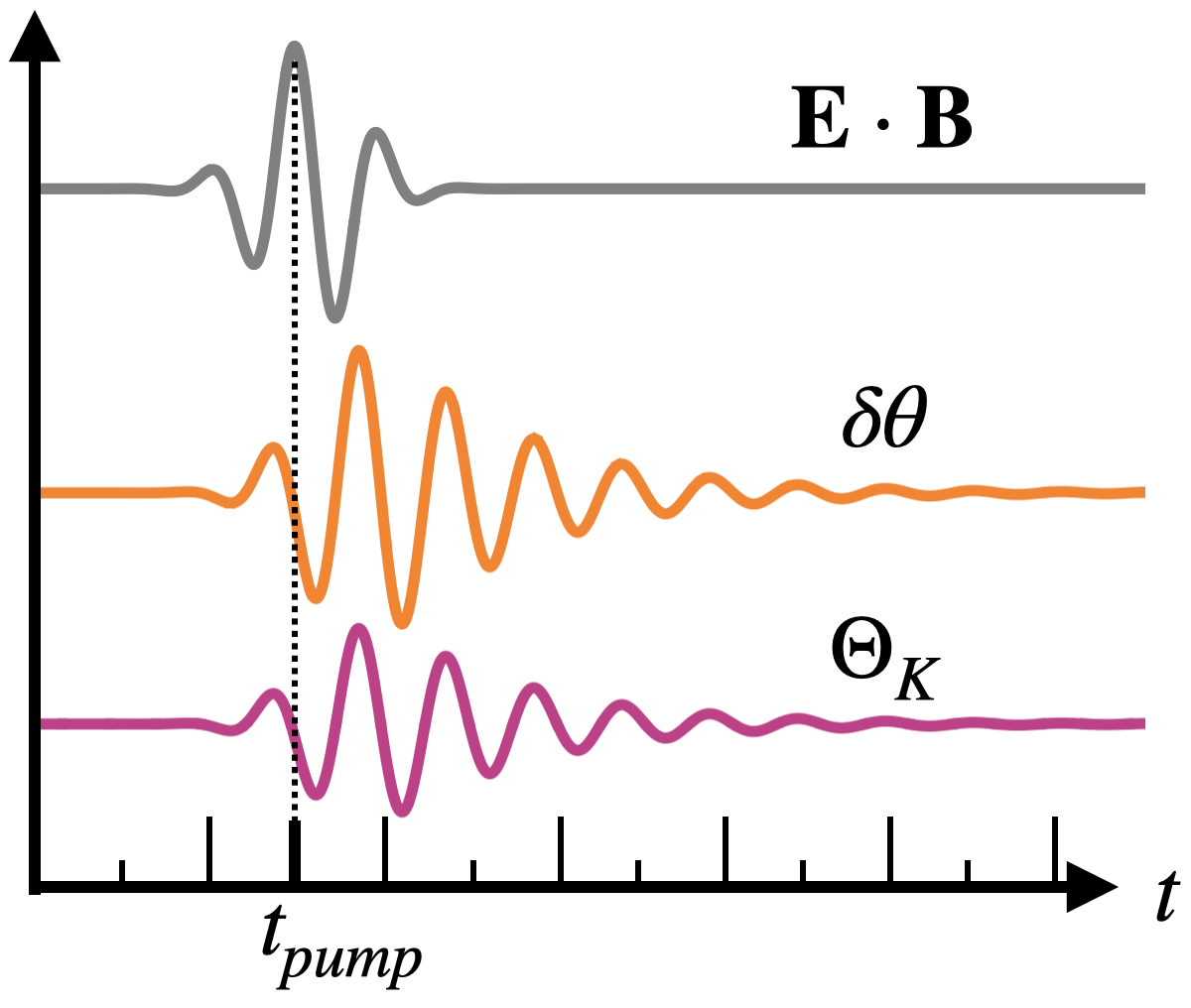}
\caption{\label{fig:AxionKerrPumpResponse} 
Transient Kerr rotation reveals signatures of excited axion modes.
Using a pump profile $\boldsymbol{E}\cdot\boldsymbol{B} \propto e^{(t/\sigma)^2} \cos(\Omega_D t)$ tuned to two-photon resonance ($\Omega_D = \Omega_0$), we drive out-of-phase axion oscillations $\delta\theta(t)$ with relative phase $\pi/2$ set by damping $\gamma$. These oscillations are reflected in the time-dependent Kerr angle $\Theta_K(t)$, which closely follows the axion dynamics. Amplitudes are illustrative and not to scale.
}
\end{figure}

The transient dynamics of the polar Kerr rotation $\Theta_K$ approximately tracking the dynamical axion field oscillations is depicted in Fig.~\ref{fig:AxionKerrPumpResponse}. 
We emphasize that the amplitudes shown are not to scale, rather this figure is meant to illustrate the coherence between $\Theta_K$ and $\delta \theta$.  
The axion equation of motion  Eq.~\eqref{eq:EOM_theta} can be understood as a driven, damped harmonic oscillator, whereby following excitation the damping term $\gamma$ causes a ring down of the axion field.
To simplify the dynamics, we take an effective force given by a Gaussian wavepacket $\boldsymbol{E}\cdot\boldsymbol{B}  \propto  e^{(t/\sigma)^2} cos(\Omega_D t) $, where $\sigma$ is the spread, and $\Omega_D$ is the driving frequency. 
As a result, the dynamical axion mode and polar Kerr rotation angle oscillate in-phase at the frequency of the drive, with the amplitude maximized when on-resonance, i.e., when $\Omega_D = \Omega_0$.

\textit{Material application}\textemdash
Up to now the treatment has been generic for any three-dimensional axionic material in the presence of strong interactions where the axion becomes a fluctuating field. 
The minimal model of such a dynamical axion insulator is defined by a set of four, anti-commuting elements, that form the basis of the four-dimensional Clifford algebra $Cl_4$, and a chiral operator~\cite{qi2008topological,petrides2022semiclassical}. 
The phenomenology of this model is equivalent to a correlated Weyl fermion, where the low-energy Hamiltonian is determined by a three-dimensional linear dispersion and a mass with two degrees of freedom, e.g., amplitude and phase.
The topology of the ground state at half-filling is characterized by an axion field, which can even become quantized when chiral symmetry is preserved. 
Within this model, the axion field is intimately related to strong interactions which can endow a dynamical behavior; specific material realizations are given by the corresponding irreducible representations.

For example, a Weyl semimetal with charge density wave order may host dynamical axion modes. In an uncorrelated Weyl semimetal with no CDW the nodes are gapless and the chirality of the Weyl nodes is given by the topological charge associated to each band crossing, which is identical to the Berry flux $C = +1$ ($C = -1$) for the right- (left-) handed Weyl particles~\cite{Yan_Binghai_Felser} for a spinful system. However, in the presence of sufficiently strong interactions, a transition to a CDW ordered phase with complex order parameter $\Delta$ can occur.

Due to the chiral anomaly, it is known that the phase of the complex CDW can be identified with the axion response~\cite{Wang.2013} of the material. The total CDW phase $\theta$ is given by $\theta(\boldsymbol{r},t) =  \bm Q\cdot \bm r + Q_0 t +  \delta\theta(\boldsymbol{r},t)\,,$
\comment{
\begin{eqnarray}\label{eq:WeylAxion}
    \theta(\boldsymbol{r},t) =  \bm Q\cdot \bm r + Q_0 t +  \delta\theta(\boldsymbol{r},t)\,,
\end{eqnarray}
}
where the first two terms are dubbed the background axion field  $\theta_0 = \bm Q\cdot \bm r + Q_0 t $ and are determined by the band structure of the material: the separation in momentum-space $\bm Q$, and energy $Q_0$, between the Weyl nodes. 
The former is non-zero only in the case of time-reversal symmetry breaking, while the latter is non-zero for inversion-symmetry breaking systems.
The final term $\delta\theta(\boldsymbol{r},t)$ is the dynamical axion mode, which can be understood as the collective mode associated to the sliding phase mode of the CDW. 
In this case, the parameter $\kappa$ in Eq.~\eqref{eq:FullLagrangianLorentzForm} is physically interpreted as the chiral compressibility.

Real material candidates include the CDW Weyl semimetal (TaSe$_4$)$_2$I~\cite{Gooth.2019}, or magnetic Weyl Co$_3$Sn$_2$S$_2$~\cite{Kanagaraj.2022}. The key difference between the non-magnetic and magnetic Weyl semimetals is the presence or absence of time-reversal symmetry, respectively, thus yielding a different form of the dielectric tensor. 
We carry out an order-of-magnitude estimate of the size of this effect for the example of the Weyl semimetal with CDW order material system (TaSe$_4$)$_2$I, with detailed material parameters given in the S.I.. The axion-induced oscillations to the polar Kerr rotation angle are estimated to be on the order of $\Theta_K \approx 1.5  \mu$rads, a small but observable response 
\textcolor{black}{by current experimental means \cite{Kapitulnik_2023, Kapitulnik_2009}. Our signal estimates, detailed in the S.I., are in fact conservative and based on less optimal geometries. More favorable configurations yield even stronger expected responses. Potential noise sources—such as thermal fluctuations, laser instability, and stray electromagnetic fields—can be mitigated with cryostats, optical choppers, and shielding, respectively. See S.I. for expanded discussion. }

\textit{Conclusion}\textemdash To conclude, we have shown that, for a generic, three-dimensional, axion insulating system with strong interactions, a two-step excitation-detection protocol may be used to unambiguously detect the presence of a dynamical axion mode in the bulk via its effects on the polar Kerr rotation angle. 
To model the dynamical axion and determine a novel optical signature we perturbatively solved the nonlinear partial differential equations of Eq.~\eqref{eq:EOM_theta} describing its dynamics. 
We established a two-step, pump-probe style protocol which excites this mode via a two-photon excitation scheme, and subsequently detects it with a third probe beam.
We showed that the intensity of the axion excitation is stronger for parallel collinear beams relative to other beam configurations, understood as the stimulated Raman scattering pathway.  
Our key result is the proposed smoking-gun optical signature, manifested in the modulation of the polar Kerr rotation which approximately  tracks the oscillations in the underlying axion field. 

\comment{
A motivation of this work is to provide theoretical predictions that can drive future experimental work, where potential material candidates we have mentioned could include axion insulators (TaSe$_4$)$_2$I \cite{Gooth.2019,Kim.2021,Nguyen.2022} and MnBi$_2$Te$_4$~\cite{Zhang.2019} (MBT). 
From a fundamental physics perspective, intuition gleaned from detecting the condensed matter axion may offer insight when mapped back to its counterpart in quantum chromodynamics. 
The condensed matter axion  
arising in topological materials allows for a more accessible, tabletop setting  for studying the more inaccessible world of high energy particle physics. 
The QCD axion, possibly within reach of experimental verification by current experiments, represents one of the most compelling pathways beyond the Standard Model, by providing at the same time an elegant solution to the strong CP problem and an excellent dark matter candidate~\cite{Day_Ellis_Marsh, Peccei_Quinn_1997}.  
Our work also has implications for condensed matter axion systems that are being used to detect their high energy counterparts in direct detection dark matter experiments ~\cite{Sch_tte_Engel_2021, Chigusa_2021}. 
}

\textcolor{black}{Definitive excitation of the dynamical axion enables control of its nonlinear magnetoelectric effects for technological use. This could advance optical technologies such as nonreciprocal thermal emitters ~\cite{Zhao.2020,otey2010thermal}, rectifiers, and infrared upconversion detectors ~\cite{Liu.2017, Chang.2007, Nastos.2006}, without relying on strong magnetic fields. In photonics, the axion’s strong electric-magnetic coupling may enable quantum logic gates, long-range quantum communication, and low-power laser systems. Additionally, two-photon excitation offers a pathway to nonlinear quantum spectroscopy and enhanced quantum sensing near the Heisenberg limit.}

\begin{acknowledgments}
The authors would like to acknowledge fruitful discussions with Aaron Chou, Fahad Mahmood, Soyeun Kim, Ankit Disa, Yikun Wang, and Emily Been.
This work is entirely supported by the Quantum Science Center (QSC), a National Quantum Information Science Research Center of the U.S. Department of Energy (DOE). P.N. gratefully acknowledges support from the Gordon and Betty Moore Foundation grant number \#8048 and from the John Simon Guggenheim Memorial Foundation (Guggenheim Fellowship).
\end{acknowledgments}

\bibliography{AxionRev5}

\newpage
\onecolumngrid
\appendix

\subsection{Derivation of axion response function}\label{apx:A}

Here we provide more detailed derivations of the results presented in the main work and some additional analysis. We begin by outlining the steps to excite the axion mode via two externally applied light beams, and arrive at the axion response function.  To start, we show how the gauge field and axion equations of motion may be derived from the full axion-modified electrodynamic Lagrangian 
\begin{eqnarray}
    && \mathcal{L} =  \frac{1}{2} ( \epsilon \boldsymbol{E}^2 - \frac{1}{\mu} \boldsymbol{B}^2 ) + \frac{\alpha}{\pi} \theta \boldsymbol{E} \cdot \boldsymbol{B} + \frac{\kappa}{2} [ (\partial_t \delta\theta)^2  - \Omega_0^2  \delta\theta^2 ] , \label{app:lagrangian1}
\end{eqnarray}
by utilizing the Euler-Lagrange formalism. To simplify things, we switch to writing in terms of potentials $\boldsymbol{E} = - \nabla \phi - \partial_t \boldsymbol{A}$ and $\boldsymbol{B} = \nabla \times \boldsymbol{A}$ where $\phi$ is the scalar potential and $\boldsymbol{A}$ is the vector potential. Plugging in for these yields 
\begin{multline}
    \mathcal{L} =  \frac{\epsilon}{2} (-\nabla \phi - \partial_t \boldsymbol{A})^2 - \frac{1}{2 \mu} (\nabla \times \boldsymbol{A})^2  + 
     \frac{\kappa}{2}\left[  (\partial_t \delta \theta)^2 - \Omega_0^2 \delta \theta^2) \right] 
    + \frac{\alpha}{\pi} \theta \left[ -\nabla \phi (\nabla \times \boldsymbol{A})  - (\partial_t \boldsymbol{A})(\nabla \times \boldsymbol{A}) \right].
\end{multline}
We choose the Weyl gauge such that $\phi = 0$. 
The Euler-Lagrange equation for $\boldsymbol{A}$ is 
\begin{equation}
    \frac{\partial }{ \partial t} \left( \frac{\partial L}{\partial (\partial_t \boldsymbol{A})} \right) + \nabla \cdot \left( \frac{\partial L}{\partial (\nabla \boldsymbol{A})} \right) - \frac{\partial L}{\partial \boldsymbol{A}} = 0.
\end{equation}
Varying with respect to $\boldsymbol{A}$ yields its equation of motion 
\begin{equation}
        \epsilon \partial^2_t \boldsymbol{A} - \frac{1}{\mu} \nabla \times \nabla \times \boldsymbol{A} =  \frac{\alpha}{\pi}  \partial_t \theta \nabla\times\boldsymbol{A}.  \label{eq:SI_A_EOM}
\end{equation}
Similarly for the axion field
\begin{equation}
    \frac{\partial }{ \partial t} \left( \frac{\partial L}{\partial (\partial_t\theta)} \right) + \nabla \cdot \left( \frac{\partial L}{\partial (\nabla \theta)} \right) - \frac{\partial L}{\partial \theta} = 0 ,
\end{equation}
and the corresponding equation of motion
\begin{equation}
     \kappa (\partial_t^2 + \gamma \partial_t + \Omega_0^2  )\delta \theta  = \frac{\alpha}{\pi} [ \partial_t \boldsymbol{A} (\nabla \times \boldsymbol{A}) ]. \label{eq:axion_EOM_SI}
\end{equation}
Terms proportional to $\nabla \delta \theta$ may be dropped by observing the speed of the axion (i.e. speed of sound) is much slower as compared to the speed of light in the medium, as such we assume the axion is dispersionless. This assumption is relaxed only when we consider the discontinuity that occurs in  $\nabla \delta \theta$ at the boundary between medium and vacuum. This is discussed in greater detail when we consider boundary conditions below.   

We now describe the protocol to excite a dynamical axion mode, where the approach is kept general and may apply to any axionic, correlated system which may endow the axion field with intrinsic dynamics. 
For ease of calculation we switch to the electromagnetic gauge field notation which relates the electric and magnetic fields $\boldsymbol{E} = - \nabla \phi - \partial_t \boldsymbol{A}$ and $\boldsymbol{B} = \nabla \times \boldsymbol{A}$ to the scalar potential $\phi$ and vector potential $\boldsymbol{A}$.  
For simplicity, we take the Weyl gauge where $\phi = 0$. 
Perturbatively expanding $\delta \theta$ and $\boldsymbol{A}$ in the small parameter $\alpha$, i.e., the fine structure constant,
\begin{align}
    & \boldsymbol{A}(\boldsymbol{x},t) = \boldsymbol{A}^{(0)} + \alpha \boldsymbol{A}^{(1)} + \alpha^2 \boldsymbol{A}^{(2)} + \alpha^3 \boldsymbol{A}^{(3)} + ... , \\
    &\delta\theta(\boldsymbol{x},t) = \delta \theta^{(0)} + \alpha \delta \theta^{(1)} + \alpha^2 \delta\theta^{(2)} + \alpha^3 \delta\theta^{(3)} + ...  .\label{eq:theta_expansion}
\end{align}
We are chiefly concerned with how the axion responds to interactions with incident light. As such, we include only relevant equations to this end. 
Plugging in the above expansions into the Eq.~\eqref{eq:axion_EOM_SI} axion equation of motion, and collecting terms according to their order in $\alpha$, we can determine the lowest surviving order. We take the constant term $ \delta \theta^{(0)} = 0$ as there are no dynamical axion excitations to start:
\begin{equation}
    \kappa (\partial_t^2 + \gamma \partial_t + \Omega_0^2 ) \delta \theta^{(1)} =  \frac{1}{\pi}  (\partial_t \boldsymbol{A}^{(0)}))(\nabla \times \boldsymbol{A}^{(0)})). \label{eq:SI_theta2}
\end{equation}
The lowest order in $\delta \theta$ is second order in the fields. Thus the nonlinear coupling of $\delta \theta$ to $\boldsymbol{A}$ is a second order process, and minimally requires two interactions with electromagnetic fields to excite. 

With this in mind, we solve for $ \delta \theta^{(1)}$ by Fourier transforming Eq.~\eqref{eq:SI_theta2}. Note the vector potential Fourier transforms as $\boldsymbol{A} (x_i) = \int_q e^{i q_i \cdot x} \boldsymbol{A}(q_i) $; and we denote $\int_q = \int \frac{d^3q}{(2 \pi)^3} \int \frac{d \Omega }{2 \pi}$, and $q \cdot x = \boldsymbol{q}\cdot \boldsymbol{x} - \Omega t$ in the exponential.  The left-hand side is
\begin{align}
    \kappa \int d^4x e^{- i q \cdot x} (\partial_t^2 +  \gamma \partial_t+ \Omega_0^2 ) \delta \theta^{(1)}(x) = \nonumber \\
    \kappa (-\Omega^2 + i \gamma \Omega + \Omega_0^2) \int d^4x e^{- i q \cdot x} \delta \theta^{(1)}(x) = \nonumber \\
   \kappa  (-\Omega^2 + i  \gamma \Omega + \Omega_0^2) \delta  \theta^{(1)}(q),
\end{align}
where $\Omega$ is the frequency of the axion, and $\Omega_0$ is its resonance. 
Inverting the matrix $(-\Omega^2 + i \Omega + \Omega_0^2) \mathbb{I}$ to isolate $\delta \theta^{(1)}(q)$ on the left-hand side: 
\begin{eqnarray}
   && =  [-\Omega^2 + i  \gamma \Omega + \Omega_0^2]^{-1} (\frac{1}{\pi \kappa}) \int d^4x e^{- i q \cdot x} [\partial_t \int e^{i q_1 \cdot x} \boldsymbol{A}^{(0)}(q_1) \cdot \nabla \times \int e^{i q_2 \cdot x} \boldsymbol{A}^{(0)}(q_2)  ] \nonumber\\
   && = [-\Omega^2 + i  \gamma \Omega + \Omega_0^2]^{-1} ( \frac{1 }{\pi \kappa}) \int_{q_1,q_2} \int_q d^4x e^{-i(q\cdot x - q_1  \cdot x - q_2 \cdot x)} (-i \Omega_1 \boldsymbol{A}^{(0)}(q_1) ) (i \boldsymbol{q_2} \times \boldsymbol{A}^{(0)}(q_2)).
\end{eqnarray}
Making use of the exponential representation as the Dirac delta function $\int d^4x e^{-i(q\cdot x - q_1  \cdot x - q_2 \cdot x)} = \delta_{q_1 + q_2, q} $ simplifies this to 
\begin{eqnarray}
        \delta \theta^{(1)}(q) = \frac{1}{\pi \kappa}\left[-\Omega^2 + i \gamma \Omega + \Omega_0^2\right]^{-1}  \int_{q_1,q_2} \delta_{q_1 + q_2, q}  \omega_1 \boldsymbol{A}^{(0)}(q_1) \cdot (\boldsymbol{q_2} \times \boldsymbol{A}^{(0)}(q_2)).\label{eq:SI_theta2_fourier}
\end{eqnarray}
Here $q = (\Omega, \boldsymbol{q})$ is the axion four-momenta, with axion frequency $\Omega$ and momentum $\boldsymbol{q}$; with resonant frequency $\Omega_0$.
The zeroth order gauge field $\boldsymbol{A}^{(0)}(q_i)$ for $i=1,2$ describes an incident electromagnetic field with four-momentum $q_i = (\omega_i, \boldsymbol{q}_i)$ for frequency $\omega_i$ and momentum $\boldsymbol{q}_i$. 

Recall the field is minimally composed of two waves whose frequencies add up to axion frequency $\Omega = \omega_1 + \omega_2$.
We consider the following plane-wave ansatz $\boldsymbol{A}^{(0)}(q) =\mathbfcal{A}_1 \delta_{q,q_1} + \mathbfcal{A}_2 \delta_{q,q_2}$ whose field amplitudes $\mathbfcal{A}_i = \mathcal{A}_i \boldsymbol{\hat{\varepsilon}}_i$ are related to the polarizations $\boldsymbol{\hat{\varepsilon}}_i$ which obey $\boldsymbol{q}_i \cdot \boldsymbol{\hat{\varepsilon}}_i = 0$. We further assume that the momenta satisfy the free-space dispersion relations so that $|\boldsymbol{q}_i | = |\omega_i|$

Using this ansatz with Eq.~\eqref{eq:SI_theta2_fourier} we can write down  the response of the dynamical axion to the application of externally applied fields $\bm A^{(1)}(q_1)$ and $\bm A^{(1)}(q_2)$
\begin{align}
       && \delta \theta(q) = \frac{(\alpha /\pi \kappa ) \omega_1 \omega_2\mathcal{A}_1  \mathcal{A}_2  }{\Omega^2 + i \Omega \gamma - \Omega_0^2}   \left( 
\frac{\boldsymbol{q}_1}{\omega_1}  -  \frac{\boldsymbol{q}_2}{\omega_2} \right) \cdot \left(\boldsymbol{\hat{\varepsilon}_1} \times \boldsymbol{\hat{\varepsilon}_2} \right), \label{eq:Appendix_theta_soln} 
\end{align}
where $\Omega = \omega_1 + \omega_2$ since the two incident beams' frequencies will add up to axion frequency $\Omega$. 

\subsection{Axion response function analysis}
The triple product which appears in Eq.~\eqref{eq:Appendix_theta_soln} is understood geometrically as the parallelpiped volume made by the polarization unit vectors,  $\hat{\varepsilon}_i$,  and the vector composed of the three-momenta $ \boldsymbol{q}_1 / \omega_1 -  \boldsymbol{q}_2 / \omega_2$.  The excitation scheme that maximizes this volume corresponds to beams with orthogonal polarizations and which are also orthogonal with respect to their relative three-momentum $\boldsymbol{q}_1 / \omega_1 -  \boldsymbol{q}_2 / \omega_2$. 

The field required to excite the axion is minimally composed of two waves, as such as we explore the efficacy of this two-photon excitation protocol for a variety of different incident beam angles $\varphi$ and frequencies $\omega_i$ in the maximally-crossed polarization geometry, i.e., we keep the polarizations of the excitation beams orthogonal $\hat{\varepsilon}_1 \cdot \hat{\varepsilon}_2 = 0$.
Depending on the incident angle $\varphi$ made by the beams, we find two distinct mechanisms that can be used to excite the dynamical axion mode. 
First, when the excitation beams are close to head-on, i.e., $\varphi = 0$, the two-photon absorption channel appears; moreover, when the excitation frequencies match the axion's frequency, i.e., $\omega_1 + \omega_2 = \Omega_0$, the amplitude of oscillations is maximized for this case. 

On the other hand, for parallel-propagating collinear beams, i.e., $\varphi = \pi/2$, this channel is entirely suppressed and instead the stimulated (inelastic) Raman excitation channel emerges.
Note in this case the relative momentum of the two beams is zero and therefore there must be a finite relative-frequency. This detailed in the main text, this beam configuration achieves the strongest axion response. 
This can be understood by again referring to  Eq.~\eqref{eq:Appendix_theta_soln}: for configurations where $\omega_1 >0$ and $\omega_2 < 0$, this will maximize the vector $\boldsymbol{q}_1/\omega_1 - \boldsymbol{q}_2/ \omega_2$ which will become a sum instead of a difference. In turn this will maximize the parallelpiped volume, as the polarization vectors $\hat{\varepsilon_i}$ are kept fixed. 

At intermediate incident angles, e.g., $\varphi = \pi/4$, both channels are active.  
Due to the dependence of the response to the geometric volume given in Eq.~\eqref{eq:Appendix_theta_soln}, the two-photon absorption channel is significantly weaker as compared to the stimulated Raman scattering channel.

\subsection{Index of refraction in the axion insulator}\label{apx:Index}
The fluctuations of the axion field are incorporated in the displacement field $\bm D = \epsilon \bm E - \frac{\alpha}{\pi} \theta \bm B$ and magnetic field strength $\bm H =  \frac{1}{\mu}\bm B +\frac{\alpha}{\pi} \theta \bm E$, where $\epsilon$ ($\mu$) is the relative permittivity (permeability). 
In addition, we assume that the probe frequency is much higher than the pinning frequency $\Omega_0$ such that the dynamical axion contributions to the magnetic conductivity can be neglected.

In this regime, the equations of motion for the electromagnetic vector field inside the material can be written in frequency and momentum space as
\begin{equation}
    \omega^2 \left(\epsilon+i\epsilon'\right) {\bm A} =k^2 \mu^{-1}\left(1-i\mu' \right) {\bm A} ,
    \label{eq:EMeom}
\end{equation}
where  $\epsilon'=\frac{\sigma^E}{ \omega}$ is the contribution due to the electric conductivity tensor, 
and $\mu'/\mu = \frac{\sigma^B}{ k}$ is the contribution due to the magnetic conductivity tensor.
Without loss of generality we assume a non-magnetic, spatially isotropic medium such that the permittivity and permeability tensors are simplified to $\epsilon =\epsilon \mathds{1}$ and $\mu^{-1} =\mathds{1}$. 
To obtain the dispersion relation of electromagnetic fields it is sufficient to consider propagation along the z-axis and solve the equations of motion in the plane orthogonal to the direction of propagation. 

In this geometry, the electric and magnetic conductivity tensors are reduced to $\sigma^E _{ij}= \varepsilon_{ij} \partial_z \theta$ and $\sigma^B_{ij}=\varepsilon_{ij} \partial_t \theta$, respectively, with $\varepsilon_{ij}$ the two-dimensional Levi-Civita tensor.
For simplicity, the contributions of such terms to the equations of motion are written as $\epsilon' = \epsilon' \sigma_y$ and $\mu' = \mu' \sigma_y$, where $\epsilon' = \partial_z \theta/\omega$, $\mu' = \partial_t \theta/k$, and $\sigma_y$ is the second Pauli matrix.
Solving Eq.~\eqref{eq:EMeom} naturally yields the chiral basis $\boldsymbol{A}_{\pm} = A_{\pm} \frac{1}{\sqrt{2}} \left(
1 ,
\pm i\right)^T$ where $A_{\pm}$ is an amplitude prefactor, with corresponding dispersion $k_{\pm}  = \omega n_\pm$ where the refractive index is given by 
\begin{eqnarray}\label{eq:Appendix_indexrefraction}
    n_{\pm}^2 = \frac{\epsilon \pm \epsilon'}{1\pm \mu' }
\end{eqnarray} 
We note that for a time-reversal and inversion symmetry preserving system, the eigenvalues are degenerate $n_+ = n_-$.

\subsection{Boundary conditions and Kerr rotation angle}\label{Appendix:BC_Kerr}

We start by considering the interface between the vacuum and material as a region that separates a trivial insulator with zero total axion field, from a nontrivial dynamical axion insulator.
We take the axion field to be a continuous function that smoothly goes from zero to its bulk value across the boundary and integrate Maxwell's equations to find the Frensel equations for the electric and magnetic fields
\begin{align}
     & \boldsymbol{E}^I +\boldsymbol{E}^R = \boldsymbol{E}^T, \\
     & \boldsymbol{B}^I + \boldsymbol{B}^R = \boldsymbol{B}^T + \frac{\alpha}{\pi}  \delta \theta \boldsymbol{E}^T.
\end{align}
Due to the onset of interactions, the dynamical axion mode $\delta\theta$ experiences a jump at the boundary and, hence, has a non-zero contribution to the Fresnel equations. 
We note that any additional contributions to the matching conditions from a nontrivial background axion field at the boundary are neglected; these can be incorporated in the current description by redefining the equilibrium position of the dynamical axion mode $\delta \theta$.

We consider light propagating in the $z$-direction with dispersion in the vacuum given by $\boldsymbol{k} = \omega \hat{z}$ and in the DAI by $\boldsymbol{k} = \omega n_{\pm} \hat{z}$, as above. 
We use the notation $\bm A_\pm = A_\pm \hat{e}_{\pm}$ where $A_\pm$ is the amplitude and $\hat{e}_{\pm}$ is the circularly polarized basis vectors.
The boundary condition  for the electric field gives the simple condition $1 + R_{\pm} = T_{\pm}$, where $ R_{\pm}={A_{\pm}^R}/{A_\pm ^I}$ are the reflection coefficients, and $ T_{\pm}={A_{\pm}^T}/{A_\pm ^I}$ are the transmission coefficients.
The boundary condition for magnetic field in the chiral basis is given by 
\begin{align}
 - \hat{z} \times \boldsymbol{A}^I_{\pm} +  \hat{z} \times \boldsymbol{A}^R_{\pm} = - n_{\pm}\hat{z} \times \boldsymbol{A}^T_{\pm} -  \frac{\alpha}{\pi} \delta \theta \boldsymbol{A}^T_{\pm}.
\end{align}
Plugging in for the vector potentials and using $\hat{z}\times \hat{e}_{\pm} = \pm i \hat{e}_{\pm}$ we obtain
\begin{align}
1 -  R_\pm = \left(n_\pm + i \frac{\alpha}{\pi} \delta \theta  \right) T_\pm .
\end{align}
Using $1 + R_{\pm} = T_{\pm}$ and solving for the reflection coefficients gives the result in the main text
\begin{equation}\label{eq:Appendix_Rpm}
    R_{\pm} = \pm \frac{1 - (n_{\pm} \mp i \frac{\alpha}{\pi} \delta \theta  )}{1 + (n_{\pm} \mp i \frac{\alpha}{\pi} 
\delta \theta )}  . 
\end{equation}

From this, the complex Kerr rotation angle $\Phi_K  = \Theta_K + i \eta_K$ may be computed, where the real part $ \Theta_K$ is termed the polar Kerr rotation, while the imaginary part $\eta_K$ is the polar Kerr ellipticity. The polar Kerr rotation relates to the argument of the reflection coefficients, $R_{\pm} = |R_{\pm} | e^{i \Delta_{\pm}} $,  as $\Theta_K = - \frac{1}{2}(\Delta_+ - \Delta_-)$, where $\Delta_{\pm}$ are the phases of the respective reflection coefficient. 
Meanwhile the polar Kerr ellipticity is related to the amplitudes as $\eta_K = (R_+ - R_-)/(R_+ + R_-)$. 

The polar Kerr ellipticity $\eta_K$ arises solely from the bulk background axion field and the associated change in the refractive index $n_\pm$, see Eq.~\eqref{eq:Appendix_indexrefraction}.
On the other hand, the polar Kerr rotation $\Theta_K$ presented here is entirely distinct from this. 
It is related to the collective fluctuations of the DAI, therefore, generating a dynamical response of the polar Kerr rotation angle.
We note that an additional nontrivial background axion field at the boundary can even lead to a quantized Kerr response.

We linearize Eq.~\eqref{eq:Appendix_Rpm} for small $\delta \theta $ and focus on the real part of the complex Kerr rotation, $\Theta_K$, given by
\begin{align}\label{eq:Kerr_appendix}
   \Theta_K = - \frac{1}{2} (\Delta_+ - \Delta_- )\approx  \frac{\alpha}{\pi} \frac{(-2 + n_+^2 + n_-^2 )}{(n_+^2 -1)(n_-^2 -1)} \delta \theta .
\end{align}

\subsection{Material application}\label{Appendix:Estimate}
So far this treatment has been material agnostic and as such is applicable to any axionic insulating system. We now consider the specific material realization in the case of a Weyl semimetal with charge density wave order.  The form of the Hamiltonian expanded near the degeneracy points is
\begin{eqnarray}
 \boldsymbol{\mathcal{H}}(\boldsymbol{k}) =\left(\begin{array}{cc}
    v_F \boldsymbol{\sigma}\cdot \boldsymbol{k} & \Delta \\
    \Bar{\Delta} & -v_F \boldsymbol{\sigma}\cdot \boldsymbol{k}
\end{array}\right),
    \label{eq:Hamiltonian}
\end{eqnarray}
where $v_F$ is the Fermi velocity, $\bm \sigma$ is the triplet of Pauli matrices, $\boldsymbol{k}$ is the crystal momentum, and $\Delta = |\Delta| e^{i\theta}$ is the CDW term and is analogous to a complex mass term, also represented in terms of the gap amplitude $|\Delta|$ and phase $\theta$.  
In an uncorrelated Weyl semimetal with no CDW the nodes are gapless and the chirality of the Weyl nodes is given by the topological charge associated to each band crossing, which is identical to the Berry flux $C = +1$ ($C = -1$) for the right- (left-) handed Weyl particles 
for a spinful system. This chiral charges can be understood as magnetic monopoles, or sources and sinks, of the Berry curvature which become singular at the nodes and act as an effective magnetic field in reciprocal space.

In the presence of sufficiently strong interactions, a transition to a CDW ordered phase with complex order parameter $\Delta$ can occur.
If the order is incommensurate, i.e., the wave vector of the CDW is not a rational multiple of the reciprocal lattice vector, by Goldstone's theorem this will have a soft sliding mode associated with the phase of the CDW order, i.e., $\theta$. 
We note that for the case of commensurate CDWs the equations of motion for the dynamical axion field~\eqref{eq:axion_EOM_SI}  are supplemented by a confining potential that modifies the spectrum of possible excitations.
In this regime, the collective modes are equivalent to a massive Goldstone boson with a finite energy at zero momentum due to the commensuration potential. 
As the excitation mechanism given by the right-hand side of Eq.~\eqref{eq:axion_EOM_SI}, i.e., $\bm E\cdot \bm B$, is independent of the particular form of the homogeneous solutions, the dynamical axion can still be excited and probed, given that its zero-momentum energy is with reach of the excitation beams' frequencies.

Due to the chiral anomaly, it is known that the phase of the complex CDW can be identified with the axion response~\cite{Wang.2013}. 
of the material. The total CDW phase $\theta$ is given by
\begin{eqnarray}\label{eq:WeylAxion}
    \theta(\boldsymbol{r},t) =  \bm Q\cdot \bm r + Q_0 t +  \delta\theta(\boldsymbol{r},t)\,,
\end{eqnarray}
where the first two terms are dubbed the background axion field  $\theta_0 = \bm Q\cdot \bm r + Q_0 t $ and are determined by the band structure of the material: the separation in momentum-space $\bm Q$, and energy $Q_0$, between the Weyl nodes. 
The former is non-zero only in the case of time-reversal symmetry breaking, while the latter is non-zero for inversion-symmetry breaking systems.
The final term $\delta\theta(\boldsymbol{r},t)$ is the dynamical axion mode, which can be understood as the collective mode associated to the sliding phase mode of the CDW. 
In this case, the parameter $\kappa$ in Eq.~\eqref{app:lagrangian1} is physically interpreted as the chiral compressibility. 

Finally, we do an order-of-magnitude estimate of the size of this effect for the example of the Weyl semimetal with CDW order material system (TaSe$_4$)$_2$I with transistion temperature $T_c = 263$K.
The pinning frequency, understood as the axion resonant frequency $\Omega_0$, for $T \ll T_c $ is on the order of $1$THz. 
The two excitation beams we take to be of same frequency, $\omega_1 = \omega_2 = 0.5$THz, in the two-photon absorption channel configuration where relative angle between the beams is $\phi = 0$. 
The magnetic field scales as $B \sim E/c$ where c is speed of light; for a large but physically realizable electric field is taken on the order of $1$MV/cm. 
As the axion is assumed to be dipsersionless its momentum is taken to be zero. The momentum of the incoming excitation photons are on the order of millimeter wavelength, which is negligibly small on the scale of collective mode dispersion.  
This is a time-reversal invariant system such that the dielectric tensor is a diagonal matrix, with index of refraction $n \approx 2$, 
assuming an $800$nm probe wavelength. 

The chiral compressibility is related to the density of states at the Fermi level, $\kappa \approx 0.0017$  states/(eV \AA$^3$); 
and the damping parameter is taken as $\gamma = 0.1$THz. 
Performing the two-step excitation-detection protocol proposed in this work on a Weyl semimetal CDW system, we let the normal incidence of the probe beam
be collinear with the static axion response $\boldsymbol{Q}$. Using these parameters we calculate the axion-induced oscillations to the polar Kerr rotation angle on the order of $\Theta_K \approx 1.5 
\mu$rads, a small but observable response.

\textcolor{black}{To the point of experimental feasibility, the authors have consulted with multiple experimental groups. The understanding gleaned from these consultations is that groups working who have done work on time-resolved Kerr are well-positioned to do this type of measurement. 
In regards to potential noise sources, these are generally expected to be no different from established pump-probe time-resolved Kerr spectroscopy noise sources, which are contingent on the experimental set-up. Some primary noise sources could be attributed to thermal fluctuations in the sample or ambient temperature fluctuations, fluctuations in laser intensity or laser polarization, or stray magnetic or electric field coupling to the sample and resulting in unintended magnetoelectric fluctuations. In regards to thermal fluctuations, as far as the temperature of the material sample is far away from its transition temperature (either into an antiferromagnetic phase for topological insulators or charge density wave phase for Weyl semimetals), this should not be an issue, particularly if the material is placed in a cryostat. Similarly the optical setup could also be temperature controlled so as to mitigate any noise which might induce a drift in measurement. For concerns regarding laser-induced noise sources, an optical chopper may be used to achieve precise frequency control and stability. Optical choppers are particularly well suited for amplification of signal-to-noise ratios for time-resolved Kerr measurements when the aim is to detect a small signal in a noisy background, and are well established for this type of experimental setup. For potential field fluctuations coupling to the axion order in the material, a straightforward mitigation technique would be shielding around the experiment to prevent stray fields. Finally, there are several works in the literature of performing time-resolved Kerr spectroscopy of quasiparticles in similar experimental setups e.g. magnon modes \cite{Kohmoto.2018}, \cite{Dzyapko.2016}. 
}

Another proposed dynamical axion insulating system is the antiferromagnetic topological insulator, such as BiFe$_2$Se$_3$
or  MnBi$_2$Te$_4$ (MBT). 
Here, the dynamical axion modes can couple to the spin-wave excitations of the AFM order parameter. 
Our approach is sufficiently generalized so as to apply to this family of materials as well, yet we keep further detailed calculations the topic of future work. 

\subsection{Outlook and application}

A motivation of this work is to provide theoretical predictions that can drive future experimental work, where potential material candidates we have mentioned could include axion insulators (TaSe$_4$)$_2$I and MnBi$_2$Te$_4$ (MBT). 
From a fundamental physics perspective, intuition gleaned from detecting the condensed matter axion may offer insight when mapped back to its counterpart in quantum chromodynamics. 
The condensed matter axion  
arising in topological materials allows for a more accessible, tabletop setting  for studying the more inaccessible world of high energy particle physics. 
The QCD axion, possibly within reach of experimental verification by current experiments, represents one of the most compelling pathways beyond the Standard Model, by providing at the same time an elegant solution to the strong CP problem and an excellent dark matter candidate.
Our work also has implications for condensed matter axion systems that are being used to detect their high energy counterparts in direct detection dark matter experiments. 

\textcolor{black}{Being able to definitely excite the dynamical axion mode, allows a path forward for using and manipulating the associated physical effects for use in technological application.
The magnetoelectric coupling mediated by the axion response and its inherent nonlinear dynamics may be of practical value in optical technologies,  e.g., in nonreciprocal thermal emitters or rectifiers~\cite{Zhao.2020,otey2010thermal}. 
This could be achieved by exploiting the strong violation of Kirchhoff's law the axion field gives rise to, without the need for strong magnetic fields.
Within photonic information processing this could include photonic quantum logic gates and error-corrected long-distance quantum communications; or other photonic technologies such as low-power parametric laser systems, and novel infrared upconversion detectors~\cite{Liu.2017, Chang.2007, Nastos.2006}.  The strong coupling between electric and magnetic fields mediated by the dynamical axion field could offer novel ways to control and interact with quantum states of matter, which may lead to new mechanisms for implementing quantum gates.
Meanwhile, in the field of quantum metrology, squeezing of quantum fluctuations for improved quantum precision measurement towards the Heisenberg limit is highly desirable. 
The two-photon excitation described in this work could be used as a means to perform nonlinear quantum spectroscopy for novel quantum sensing applications.}

\subsection{Comparison to existing detection techniques}\label{Appendix:TechniqueComparison}
 \textcolor{black}{Our all-optical, zero-field approach contrasts with earlier proposals. Ahn et al. [Nat. Comm. 2022] focus on static, frequency-dependent axions requiring TRS breaking in magnetic TIs, while we study a general dynamical axion without such symmetry constraints. Li et al. [Nat. Phys. 2010] use torque magnetometry and strong external fields to linearize the axion dynamics; in contrast, we solve the full nonlinear response optically. Gooth et al. [Nature 2019] rely on transport and observe current responses attributed to axions, though replication attempts have questioned the origin of their signal. Our Kerr-based protocol directly probes the axion field and avoids thermal artifacts. Smith et al. [arXiv:2411.03432] propose a linearized optical approach, whereas we retain nonlinear dynamics to access richer physics.
 }

\end{document}